# Combining freeform optics and curved detectors for wide field imaging: a polynomial approach over squared aperture

EDUARD MUSLIMOV,[1,2,*] EMMANUEL HUGOT,[1] WILFRIED JAHN,[1] SEBASTIEN VIVES,[1] MARC FERRARI,[1] BERTRAND CHAMBION,[3] DAVID HENRY[3] AND CHRISTOPHE GASCHET[1,3]

[1]*Aix Marseille Univ, CNRS, LAM, Laboratoire d'Astrophysique de Marseille, 38, rue Joliot-Curie, Marseille,13388, France*
[2]*Kazan National Research Technical University named after A.N. Tupolev –KAI, 10 K. Marx, Kazan 420111, Russian Federation*
[3]*Univ. Grenoble Alpes, CEA, LETI, MINATEC campus, F38054 Grenoble, France*

*\*eduard.muslimov@lam.fr*

**Abstract:**. In the recent years a significant progress was achieved in the field of design and fabrication of optical systems based on freeform optical surfaces. They provide a possibility to build fast, wide-angle and high-resolution systems, which are very compact and free of obscuration. However, the field of freeform surfaces design techniques still remains underexplored. In the present paper we use the mathematical apparatus of orthogonal polynomials defined over a square aperture, which was developed before for the tasks of wavefront reconstruction, to describe shape of a mirror surface. Two cases, namely Legendre polynomials and generalization of the Zernike polynomials on a square, are considered. The potential advantages of these polynomials sets are demonstrated on example of a three-mirror unobscured telescope with F/#=2.5 and FoV=7.2x7.2°. In addition, we discuss possibility of use of curved detectors in such a design.

## **1. Introduction**

Freeform optical surfaces open a new window for optical design of innovative systems. There is a constantly growing number of examples demonstrating that use of freeform surfaces allows to build imaging systems with improved resolution across an increased field of view together with a larger aperture. These examples spread over different application fields and use different approaches to describe the freeform surfaces.

In the current study we focus on all-reflective unobscured telescope systems, which utilize freeform mirrors. We also put a special accent on applications requiring high performance like astronomy and astrophysics. The latter condition implies that we put more accent on solutions with medium-to-high apertures and high resolution.

Comparing the existing designs in the specified area, one can see that the majority of the published solutions can be classified as unobscured TMA-type telescopes, while other types like derivatives from Cassegrain or Schwarzschild designs with two mirrors or SEAL-type designs with four mirrors are rarer.

In the considered examples a few mathematical descriptions of the freeform surfaces are applied: ordinary XY-polynomials [1-7], Zernike polynomials [8, 9], non-uniform rational B-splines (NURBS) [10-12] and a few other types of equations, used in [13-17]. Among the latter it is reasonable to note Bernstein polynomials [13] and Legendre/Q-Legendre polynomials [14]. This overview is not absolutely exhaustive, but it represents the variety of existing approaches to freeform design.

To estimate and compare performance of the existing designs we unified their values of the field of view, aperture and optical quality and brought them to the charts presented on Fig. 1. Even though these diagrams neglect some important parameters like the working wavelength, they provide a rough visual estimation of the performance.

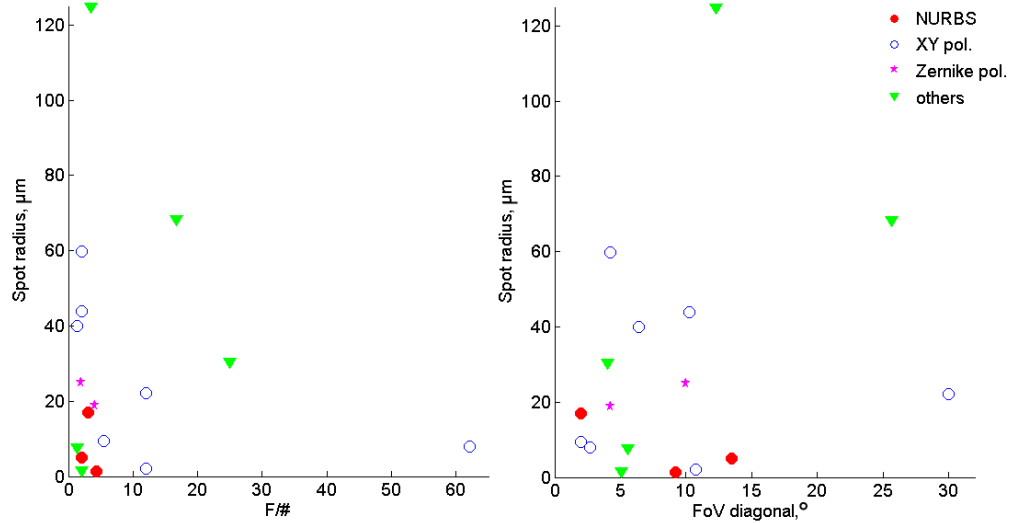

Fig. 1. Main optical parameters of existing freeform-based optical systems performance evaluation.

The diagrams show that there is no evidence of a significant priority of one of the used equations types. Moreover, each of them has some disadvantages or features which limits their application. For instance, ordinary XY-polynomials represents simple and easily available tool to model freeform surfaces, which is included in many commercial optical design softwares. But they do not compose an orthogonal basis and are not normalized, that implies that the individual terms do not have any physical meaning and tend to compensate each other during optimization in a similar way to that shown, for example, by Yongworth and Betenesky in [18] for axisymmetric aspheres. On the other hand, Zernike polynomials are widely known as an orthonormal function basis, genetically connected with wavefront aberrations. However, the classical Zernike polynomials are defined over a circular aperture, while in a majority of the covered examples the field of view is rectangular so the mirrors set remote of the pupil plane have rectangular clear aperture. The shortcomings inherent for these two common design instruments are partially removed by use of specialized polynomial bases [13,14]. Nevertheless, they were applied for specific individual design tasks and haven't become widely-accepted design tools yet.

For instance, it was shown that use of Q-Legendre polynomials allows to cover a huge field of view up to 51.4° [19], but the values of aperture and resolution achieved there makes the design incompatible with the most of optical systems used in astronomy. Finally, the NURBS-based designs shown some promising results, but they are based on a finite grid of points, that complicates their optimization with existing means. Thus we can conclude that in spite of fast progress of the freeform optics the questions of their mathematical description still remains underexplored.

Thus it is very desirable to describe freeform surfaces by such an equation, which would be defined over a square aperture, easily realizable in current optical design software and consistent with existing optimization techniques. The primary goal of this paper is to find such sets of equations and demonstrate their advantages for design of a high-aperture wide-field unobscured reflective telescope.

In the field of wavefront reconstruction and analysis similar problem was solved in a few different ways and this adjacent field can propose a number of mathematical approaches to the surface definition. Paper [20] describes use of Legendre polynomials and Zernike polynomials generalized for a case of square aperture for wavefront reconstruction. In the present study we consider possibility of their application to design of freeform mirrors. We briefly discuss their features and transform them for use in a ray-tracing code. Further we apply them for design of a fast wide-field TMA-type unobscured telescope and analyze the found design solutions.

## 2. The polynomial bases

In the current study we use the same definition of the 2D Legendre polynomials as that used in [20]. They are orthonormal over the unit square aperture and can be easily derived from 1D Legendre polynomials. However, we consider to be useful to present here a table of the polynomials written in explicit from. Firstly, it is convenient for composing a code for ray-tracing, and secondly, it is necessary for making cross-check with ordinary XY-polynomials, when performing tests. The polynomials up to the $5^{th}$ order (i.e. $21^{st}$ mode) are listed in the Table 1.

In the same way we describe the Zernike polynomials generalized for a case of a square aperture. Similarly, we use the equations proposed before for wavefront reconstruction, though in contrast with the original paper we derive all the polynomials up for the $21^{st}$ mode using the generating procedure, described in [20-22]. The explicit equations for the square Zernike polynomials practically convenient for describing a user-defined surface in an optical design software are also presented in Table 1. The given polynomials are orthonormal as well. However, we can note here, that because of a recurrent generating formalism high-order polynomials are difficult to use.

The modes of the Legendre and square Zernike polynomials sets are visualized as false color plots on Fig. 2a and 2b, respectively.

The two polynomial bases described above were used for description of freeform mirror surfaces. They were implemented as dll-files and applied for raytracing and numerical optimization in Zemax.

Table 1. Explicit form of the 2D Legendre and square Zernike polynomials

| Mode number, k | $L_k(x,y)$ | $S_k(x,y)$ |
|---|---|---|
| 1 | 1 | 1 |
| 2 | $1.7321 \cdot x$ | $1.7321 \cdot x$ |
| 3 | $1.7321 \cdot y$ | $1.7321 \cdot y$ |
| 4 | $3.3541 \cdot x^2 - 1.1180$ | $2.3717 \cdot x^2 + 2.3717 \cdot y^2 - 1.5811$ |
| 5 | $3 \cdot xy$ | $3 \cdot xy$ |
| 6 | $3.3541 \cdot y^2 - 1.1180$ | $2.3717 \cdot x^2 - 2.3717 \cdot y^2$ |
| 7 | $6.6144 \cdot x^3 - 3.9686 \cdot x$ | $4.3649 \cdot x^2 y + 4.3649 \cdot y^3 - 4.0739 \cdot y$ |
| 8 | $5.8095 \cdot x^2 y - 1.9365 \cdot y$ | $4.3649 \cdot x^3 + 4.3649 \cdot xy^2 - 4.0739 \cdot x$ |
| 9 | $5.8095 \cdot xy^2 - 1.9365 \cdot x$ | $3.8337 \cdot x^2 y - 4.9697 \cdot y^3 + 1.7039 \cdot y$ |
| 10 | $6.6144 \cdot y^3 - 3.9686 \cdot y$ | $4.9697 \cdot x^3 - 3.8337 \cdot xy^2 - 1.7039 \cdot x$ |
| 11 | $13.125 \cdot x^4 - 11.25 \cdot x^2 + 1.125$ | $4.8104 \cdot x^4 + 9.6208 \cdot x^2 y^2 + 4.8104 \cdot y^4 - 7.3302 \cdot x^2 - 7.3302 \cdot y^2 + 1.8936$ |
| 12 | $11.4564 \cdot x^3 y - 6.8739 \cdot xy$ | $9.2808 \cdot x^4 - 7.9550 \cdot x^2 - 9.2808 \cdot y^4 + 7.9550 \cdot y^2$ |
| 13 | $11.25 \cdot x^2 y^2 - 3.75 \cdot x^2 - 3.75 \cdot y^2 + 1.25$ | $8.1009 \cdot x^3 y + 8.1009 \cdot xy^3 - 9.7211 \cdot xy$ |
| 14 | $11.4564 \cdot xy^3 - 6.8739 \cdot xy$ | $7.9368 \cdot x^4 - 5.8311 \cdot x^2 y^2 + 7.9368 \cdot y^4 - 4.8593 \cdot x^2 - 4.8593 \cdot y^2 + 0.7127$ |
| 15 | $13.125 \cdot x^4 - 11.25 \cdot x^2 + 1.125$ | $8.1009 \cdot x^3 y - 8.1009 \cdot xy^3$ |
| 16 | $26.1184 x^5 - 29.0205 x^3 + 6.2187 x$ | $-18.5336 \cdot x^4 - 11.2976 \cdot y^4 - 35.1004 \cdot x^2 y^2 + 19.1136 \cdot x^2 + 13.9594 \cdot y^2 + 6.5222 \cdot x^5 + 13.0443 \cdot x^3 y^2 + 6.5222 \cdot xy^4 - 7.8266 \cdot xy^2 - 7.8266 \cdot x^3 + 1.9567 \cdot x - 3.0410$ |
| 17 | $22.7332 \cdot x^4 y - 19.4856 \cdot x^2 y + 1.9486 \cdot y$ | $16.9179 \cdot xy - 19.1009 \cdot x^3 y - 24.2833 \cdot xy^3 - 5.1293 \cdot y^5 + 10.2585 \cdot x^2 y^3 + 5.1293 \cdot x^4 y - 6.1551 \cdot y^3 - 6.1551 \cdot x^2 y + 1.5388 \cdot y$ |
| 18 | $22.1853 \cdot x^3 y^2 - 7.3951 \cdot x^3 - 13.3112 \cdot xy^2 + 4.4371 \cdot x$ | $-16.0381 \cdot x^4 + 16.6285 \cdot y^4 + 8.3475 \cdot x^2 y^2 + 11.5745 \cdot x^2 - 12.7060 \cdot y^2 + 9.4550 \cdot x^5 - 0.1453 \cdot x^3 y^2 - 9.6003 \cdot xy^4 + 5.8038 \cdot xy^2 - 9.4405 \cdot x^3 + 1.4074 \cdot x - 0.3713$ |
| 19 | $22.1853 \cdot x^2 y^3 - 7.3951 \cdot y^3 - 13.3112 \cdot x^2 y + 4.4371 \cdot y$ | $10.9852 \cdot xy - 22.10429 \cdot x^3 y - 5.4609 \cdot xy^3 - 4.9216 \cdot y^5 + 3.9945 \cdot x^2 y^3 + 8.9161 \cdot x^4 y + 4.5222 \cdot y^3 - 6.5480 \cdot x^2 y - 0.4387 \cdot y$ |
| 20 | $22.7332 \cdot xy^4 - 19.4856 \cdot xy^2 + 1.9486 \cdot x$ | $-6.7585 \cdot x^4 - 10.4893 \cdot y^4 + 10.4171 \cdot x^2 y^2 + 4.8521 \cdot x^2 + 5.1792 \cdot y^2 + 7.5073 \cdot x^5 - 6.8079 \cdot x^3 y^2 + 6.0569 \cdot xy^4 - 1.5917 \cdot xy^2 - 6.8265 \cdot x^3 + 1.3795 \cdot x - 1.0772$ |
| 21 | $26.1184 \cdot y^5 - 29.0205 \cdot y^3 + 6.2187 \cdot y$ | $-1.0178 \cdot xy - 7.5887 \cdot x^3 y + 11.6480 \cdot xy^3 \; 2.3698 \cdot y^5 - 6.2624 \cdot x^2 y^3 + 6.8826 \cdot x^4 y -1.7436 \cdot y^3 - 2.2508 \cdot x^2 y + 0.4676 \cdot y$ |

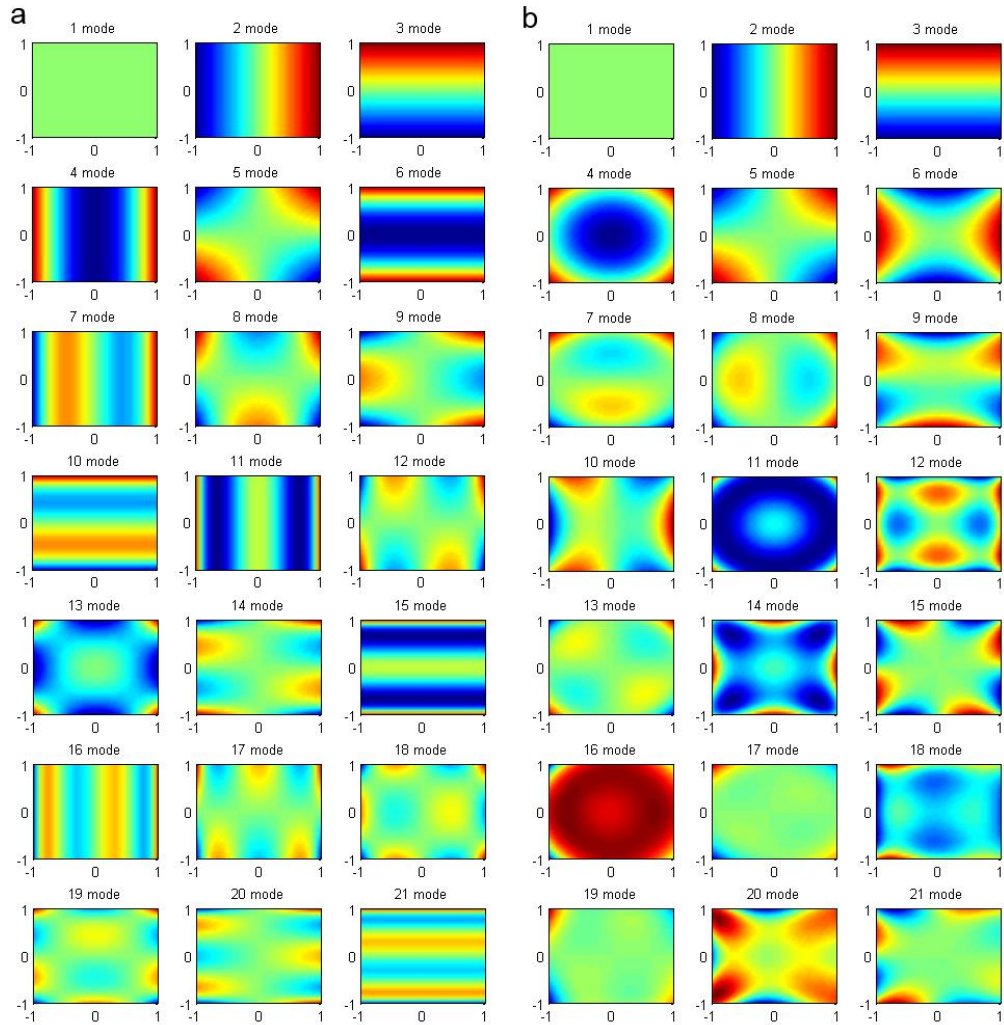

Fig. 2. Polynomials modes defined over a unite square aperture: a – Legendre polynomials, b – square Zernike polynomials.

## 3. Optical design of a TMA telescope

As an example we consider an all-reflective unobscured TMA telescope, similar to that presented in [8, 23]. We assume that all the mirrors have freeform surfaces. In contrast with the majority of analogous designs we consider cases when the detector has a curved surface. The general parameters of the optical scheme are listed in Table 2.

**Table 2. General parameters of the telescope**

| Parameter | Value | Unit |
| --- | --- | --- |
| Effective focal length | 250 | mm |
| Entrance pupil diameter | 100 | mm |
| Field of view | 7.2 x 7.2 | ° |
| Reference wavelength | 550 | nm |

We should note here that these numbers are rather demonstrative, even though they can be correlated with requirements of some prospective astrophysical mission s [9]. In addition, further significant increase of the FoV and/or F/# would make an unobscured design geometrically impossible.

The general view of the optical scheme is presented on Fig. 3. The components arrangement remains approximately the same for all the design versions discussed below.

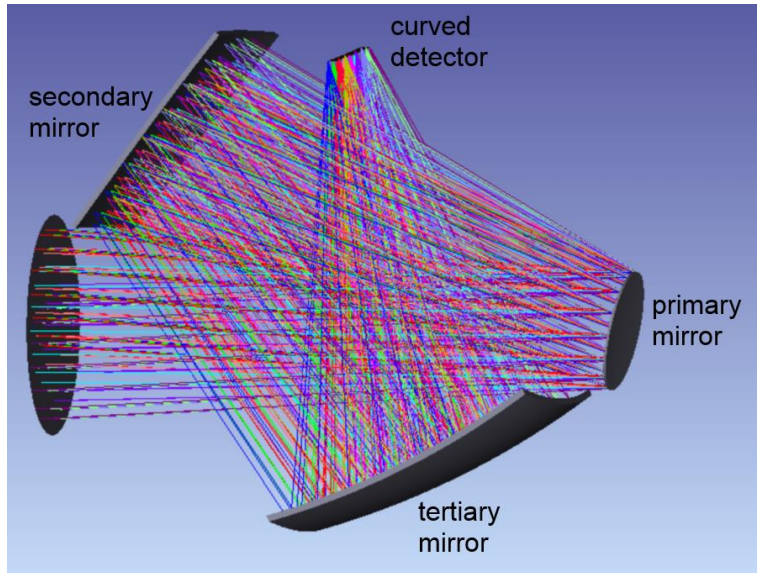

Fig. 3. Layout of the freeform-based unobscured TMA telescope.

A few design options were considered with this layout. We tested the Legendre and the square Zernike polynomials. In addition, for each of the equations types 3 options of the detector shape were explored, namely, plane, spherical and toroidal shapes. In the both of polynomials types the best results were achieved with a toroidal detector surface. Therefore, all the detailed design analysis presented further are related to this case. The impact of the detector shape on the image quality is discussed in a separate section. All the mirrors' surfaces are supposed to be symmetrical in respect to the Y axis, so the non-symmetrical terms are not used.

During the optimization the following boundary conditions were applied. Firstly, the principal rays' coordinates on the image surface are fixed to keep the proper focal length and image format. Secondly, the principal ray centering is controlled for the secondary and tertiary mirrors as well as for the detector. Further, angles of incidence and global coordinates of the marginal rays are limited in order to avoid obscuration. The sag for the center of each mirror must be equal to 0. Finally, the clear semi-apertures of the mirrors should be less than the specified normalization radii both to limit the total volume and to avoid an extrapolation.

For estimation and comparison of the image quality we use spot diagrams and energy concentration plots. The results for the telescope with mirrors described by Legendre polynomials are shown on Fig. 4, and those for square Zernike polynomials are on Fig. 5. The extreme values for comparison are summarized in Table 3. The Airy disk radius for the telescope is 1.67µm.

Table 3. Image quality indicators

| Parameter | Legendre | | Square Zernike | |
|---|---|---|---|---|
| | min | max | min | max |
| RMS spot radius, μm | 1.4 | 5.9 | 5.8 | 9.1 |
| Circumcircle radius, μm | 2.3 | 10.9 | 8.6 | 15.9 |
| Energy concentration in 5μm radius, % | 85.3 | 93.8 | 27.2 | 88.7 |

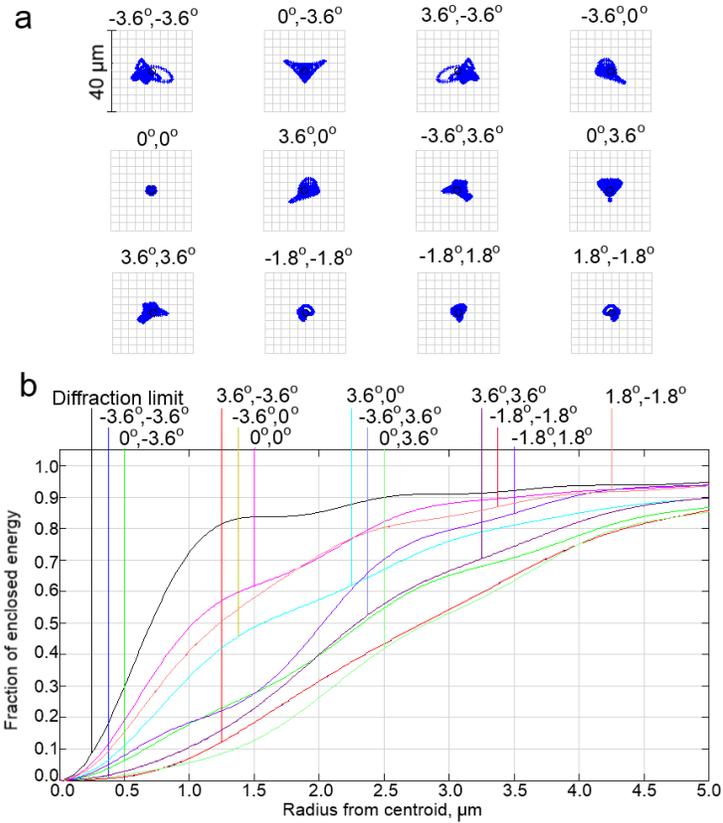

Fig. 4. Image quality of the telescope using mirrors described by Legendre polynomials: a – spot diagrams, b – energy concentration in 10μm-size pixel. The labels correspond to angular coordinates of the FoV points.

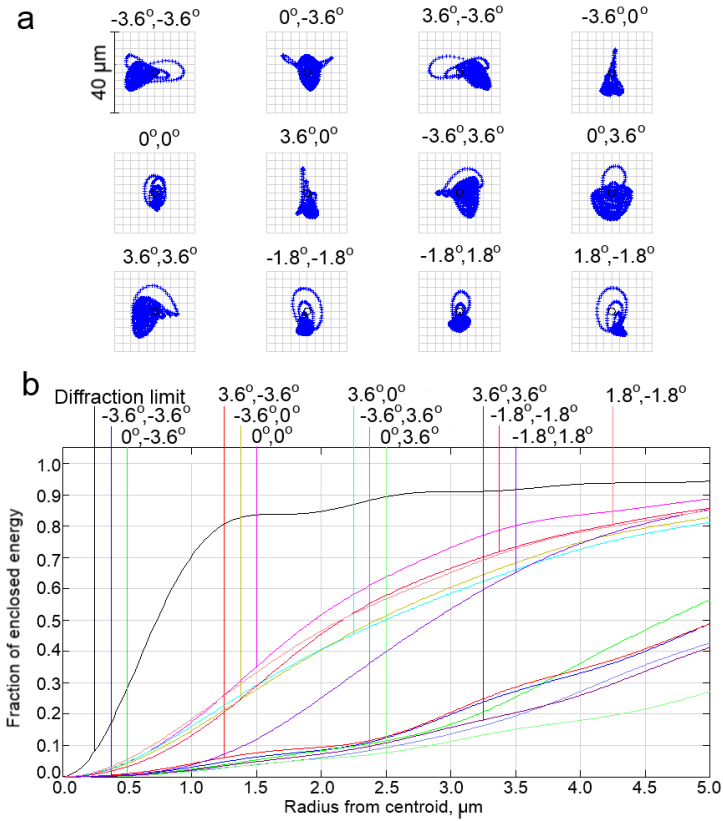

Fig. 5. Image quality of the telescope using mirrors described by square Zernike polynomials:
a – spot diagrams, b – energy concentration in 10μm-size pixel. The labels correspond to
angular coordinates of the FoV points.

The diagrams clearly show that both of the polynomials types provides relatively good aberration correction. However, the optical quality achieved with the Legendre polynomials is obviously superior over that for the square Zernike polynomials. It approaches to the diffraction limit for the central point and remains relatively uniform across all the wide 10°-diagonal field.

In addition to the image quality estimations shown above, it's necessary to mention that the design with Legendre polynomials has lower distortion (see Table 4). We can note that the design with Legendre polynomials has some advantage. Also it should be noted that the measured values are defined not only by a pure image distortion, but also by its' residual shift, which can be seen from the signs.

**Table 4. Distortion of the telescopes in %**

| Measurement direction | Legendre | Square Zernike |
|---|---|---|
| X axis, positive | -0.20 | -0.19 |
| X axis, negative | -0.20 | -0.19 |
| Y axis, positive | -0.25 | -0.31 |
| Y axis, negative | -0.24 | -0.15 |
| Vectorial | 1.33 | 1.38 |

Finally, both of the presented designs are notable for their compactness. Total volume of the telescope with Legendre-type freeforms is approximately 233x416x507 mm, while that for the design with square Zernike polynomials is 226x406x514 mm. Exact values will depend on the mirrors thicknesses and mechanical design.

## 4. Analysis of the mirrors surfaces shapes

To assess complexity of the designed freeform surfaces we use difference between the calculated surface sag and the best fit sphere (BFS). Here we define the BFS as a sphere with variable radius and center position, which provides the smallest RMS error for the current surface. It's necessary to emphasize here that the BFS provides only an approximate estimation for complexity and manufacturability of the freeforms, because in many cases [24] the surface can be grinded directly into asphere, so a spherical blank is never actually used.

All the BFS parameters and residual errors for the two designs under consideration are given in Table 5, while the residuals plots are shown on Fig. 6 for the Legendre polynomials and on Fig. 7 for the square Zernike polynomials. The units on the plots are microns. Note, that the plots are given for the full surfaces and do not take into account the real rectangular apertures.

Table 5. BFS parameters for the freeform mirrors

| | Mirror | BFS radius, mm | BFS center position, μm | RMS deviation, μm | Max. deviation, μm |
|---|---|---|---|---|---|
| Legendre | Primary | 1825.18 | 0.7 | 34.1 | 105.5 |
| | Secondary | 2768.57 | 8.0 | 251.3 | 1098.9 |
| | Tertiary | -728.27 | 6.4 | 90.7 | 354.9 |
| Square Zernike | Primary | 1444.57 | 0.7 | 34.8 | 128.1 |
| | Secondary | 1807.30 | 4.6 | 249.2 | 1053.4 |
| | Tertiary | -796.04 | 4.5 | 93.4 | 298.2 |

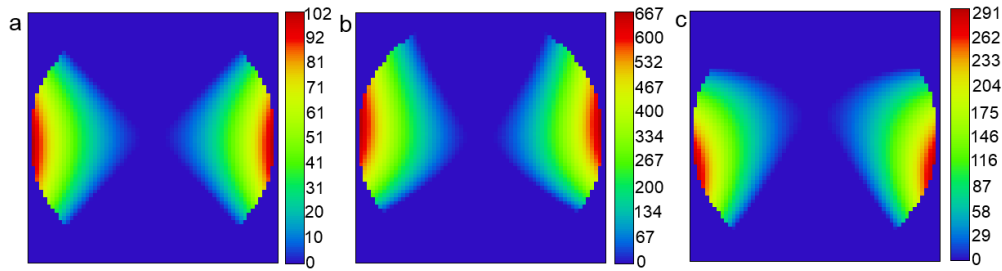

Fig. 6. Residuals after subtraction of BFS from the mirrors surfaces in the design with Legendre polynomials: a – primary mirror, b – secondary mirror, c – tertiary mirror. The units are microns (the units on colorbars are microns).

The plots indicate that the all freeform mirrors are highly anamorphic, which is expectable because of the necessity of astigmatism correction. As well as from the point of optical quality the design with Legengre polynomials has visual advantages. At least the primary and tertiary mirrors are within current manufacturing limits described Hentschel et al in [25]. The secondary mirror can bring some difficulties, though its' testability and manufacturability needs further investigation.

The design with the square Zernike polynomials has mirrors, which are approximately equivalent to that in the previous design. However, the tertiary has an uncompensated tilt equal to 5.87°. It is accounted for in the BFS computations, but during the optimization and analysis it can not be directly removed, thus complicating the design process. This disadvantage is inherent to the square Zernike polynomials and connected with the recurrent generating formulas.

Finally, it can be noted here that the tertiary mirror in both designs has the biggest optical power. This feature is common for all the optical schemes considered during the study. That propety can be useful for choice of an initial point for future designs.

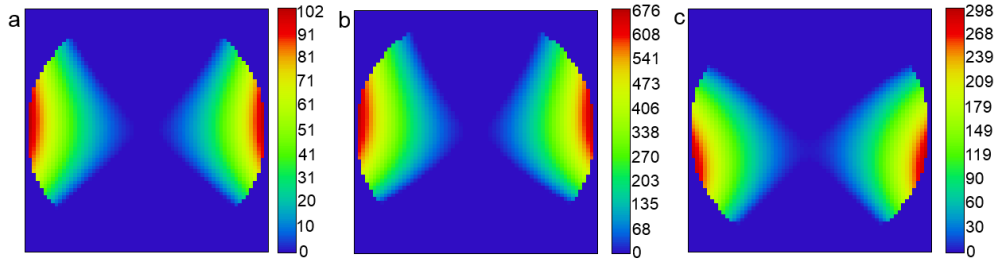

Fig. 7. Residuals after subtraction of BFS from the mirrors surfaces in the design with square Zernike polynomials: a – primary mirror, b – secondary mirror, c – tertiary mirror (the units on colorbars are microns).

## 5. Impact of the detector shape

As it was mentioned in the beginning all the detailed analysis above is related to the case when the detector has a toroidal surface. In the first design with the Legendre polynomials the tangential and sagittal radii are $R_T$=-383.73mm and $R_S$=592.95 mm, respectively. The similar values for the design with square Zernike polynomials are $R_T$=-422.70mm, $R_S$=-745.79 mm. To exemplify the detector shape the sag for the first design is plotted on Fig. 8 (the units are microns, the circle radius is 44.6 mm).

To define the impact of the detector shape on the optical quality both of the telescope designs were re-optimized, at first, for a spherical detector shape, and further for a plane shape. The results comparison is presented on Fig. 9. The bar chart represents the minimum and maximum values of the RMS spot radii for three possible shapes. The bar chart represents the minimum and maximum values of the RMS spot radii for three possible detector shapes in the schemes with Legendre and square Zernike polynomials. The spherical detector radius is -473.22mm for the first design and -534.20mm for that with the second one.

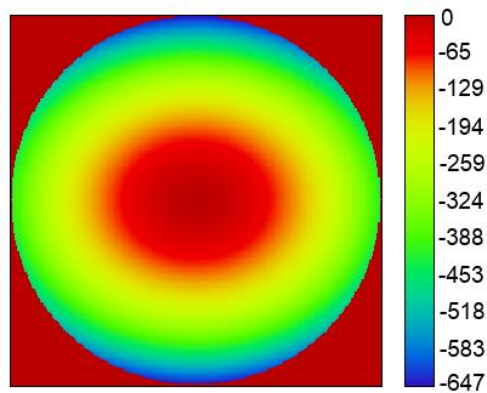

Fig. 8. Sag of the curved detector used in the design with Legendre polynomials (the units on colorbar are microns)

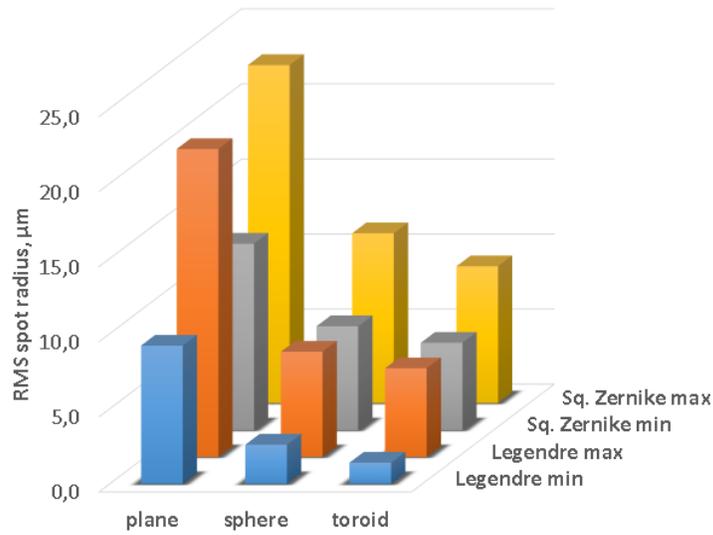

Fig. 9. Estimation of the detector shape influence on the telescopes image quality.

The chart demonstrates that the detector shape has a critical influence on the image quality. In the case of plane detector surface the spot size increases up to for 6.5 times, thus nullifying all the aberration correction achieved with the use of freeform optics. On the other hand, the difference between the spherical and toroidal detectors is not so big, but it is still significant and present regardless of the freeforms definition used. So the toroidal detector surface doesn't need a fundamentally new technology to be implemented, but it can significantly improve the telescope performance.

## 6. Notes on the detector shape feasibility

As shown in this paper, curved detectors allow a drastic simplification of the optical systems by offering a new parameter in the optimization process. The field curvature aberration, composed mainly by the Petzval curvature and astigmatism aberration, can be compensated directly in the focal plane and get rid of field-flattening optics.

Several developments of curved detectors have been pursued in recent years by Sarnoff [26], Stanford [27, 28], University of Arizona [29] and JPL [30]. Different techniques for the bending of CCD and CMOS detectors have been proposed and the prototypes give promising results, with a very low performance loss in terms of dark current, noise, and a filling factor above 80%. In astronomy, R&D activities have been undertaken by ESO for the development of large format VIS highly curved detectors [31]. In that case, the manufacturing process has been fully developed and delivered a perfectly working component suitable for astronomical applications.

The Laboratoire d'Astrophysique de Marseille (LAM) and the CEA LETI recently proposed to make use of active mirrors technology combined to flexible dies to generate deformable detectors [32]. The active mirror technology available for this application allows to generate either variable curvature detectors, variable astigmatic detectors (e.g. toroids) or even more, multi-mode detectors. The variable curvature mirror (VCM) technology consists in a circular membrane with a variable thickness distribution. A force F applied at the center of the structure and perpendicularly to the membrane surface results in an accurate spherical bending of the top surface of the membrane with a radius of curvature depending on the force F. The thinned CMOS sensor attached on top of the VCM surface follows this curvature variation (convex to concave) [32]. Finite Element Analysis have been done to calculate the VCM structure and the detector behavior during the warping. For our solution with a toroidal detector presented on Fig. 8, we combine the variable curvature technology with the generation of astigmatism as presented in [34] to get the required toroidal shape. According to Figure 14 in [33], we are able to curve a 30x30 mm size detector thinned at 100μm up to a radius of curvature of 140 mm. A thickness of 100μm is the typical the value we use for our current developments. The toroidal detector presents a tangential and sagittal radii of $R_T$=-383.73mm and $R_S$=592.95 mm, so 2.7 and 4.2 times more than the limitation value, making it safe to produce. Another way of bending would be to use tunable active mirror technology [35] developed for space mission on which the thinned CMOS sensor would be attached. This solution would allow to generate several different shapes.

There are other recently presented versions of the curved sensors manufacturing technology, for instance the pneumatic forming process described in [36].

## 7. Comparative study

The detailed description of the optical system given above should be complemented with a few notes on its' optimization process. Obviously, the obtained equations for the mirrors surfaces shapes can be rewritten with different sets of polynomials. However, not all of them are equally consistent with existing techniques and codes used for modelling and optimization of optical systems. To exemplify the difference in optimization introduced by various polynomial sets we provide we compare progress of optical quality achieved with Legendre, square Zernike, standard Zernike and XY polynomials. In each case the starting point is an unobscured system with three spherical mirrors and flat detector, which has F/# of 10 and FoV equal to 1.4°. Then the surfaces are re-defined: the spheres are turned into freeforms with zero coefficients and the detector plane is replaced by a toroidal with zero curvatures. The optical system passes through a few steps of optimization with gradually growing values of aperture and FoV. The optimization is performed by means of the standard damped least squares method [37], which is implemented in Zemax. All the parameters defining shapes and positions of mirrors and detector surfaces are used for optimization. The boundary conditions include only control of the principal rays' coordinates on the mirrors and the detector to maintain centering of the clear apertures and proper focal length value, respectively. Boundaries for the components positions are used to exclude obscuration and geometrical conflicts. There are also limits applied to the mirrors diameters. Finally, the sag value for each mirror should equal to 0 in its' center. The results obtained are shown on Fig. 10. To guide the eye, the plots are complement with a polygon showing the parameters area covered by existing designs overviewed in the introduction.

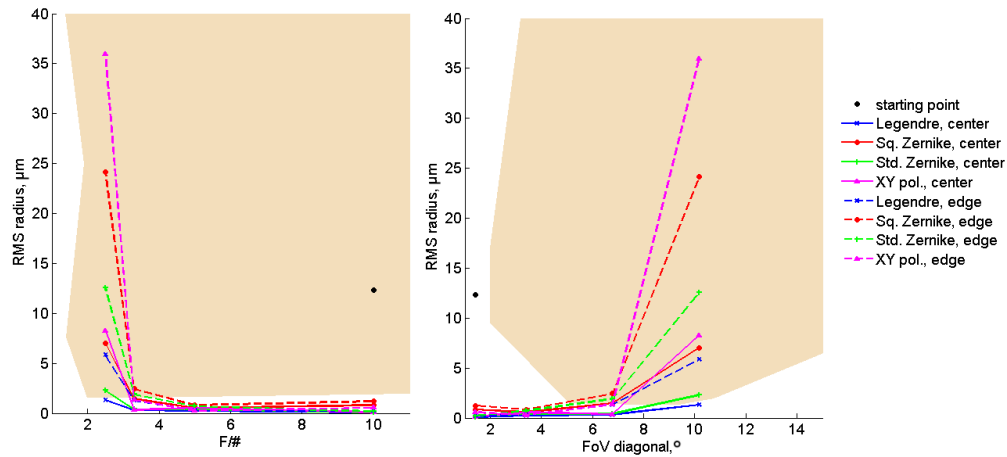

Fig. 10. Comparison of optimization results with use of different polynomial sets.

Thus, relying on the graphs we can conclude that with use of identical optimization procedure Legendre polynomials allow to achieve better performance than other types of equations. The non-normalized XY polynomials show the worst performance as it was anticipated. The standard Zernike polynomials show notable decrease odf resolution for the FoV edge. Possible explanation is the mismatch of clear apertures and circular basis, which is hard to account for using standard optimization tools. Large spot radii values, obtained with the square Zernike polynomials can represent a consequence of excessive complexity of the high-order polynomials and the necessity to compensate for the intercept. Finally, it can be noted that the image quality obtained for the FoV center with Legendre polynomials is slightly better than for existing systems with the same F/# and FoV.

## 8. Conclusion

We considered possibility of use of two orthonormal sets of polynomials for design of freeform mirrors, namely the Legendre polynomials and square Zernike polynomials. On an example of unobscured all-reflective TMA-type telescope we demonstrated that application of such polynomials for description of optical surfaces allows to achieve high performance for F/# value up to 2.5 and over a large 10°-diagonal field of view.

The Legendre polynomials shows better results in terms of geometric aberrations and energy concentration. The difference in the latter parameter for the field edge is as high as 3 times in comparison with the square Zernike polynomials. The mirrors surfaces designed with the both of the polynomials sets are within the current manufacturability limits.

Finally, it was found that use of a toroidal shape of the detector surface can improve the telescope performance significantly.

As our comparative study has demonstrated, use of the Legendre polynomials allows to achieve very high performance in comparison with other types of equations, even if a very simple starting point and only standard optimization tools are used.

In our opinion the use of the two considered sets of polynomials, especially the Legendre's ones, is a natural way to design freeform mirrors with square apertures. It can be easily implemented in practice and matches with the standard optimization tools and techniques, currently available for optical designers. We hope that the presented approach can become a common one replacing use of ordinary XY-polynomials and suggesting an alternative to the standard Zernike polynomials.

The developed telescope design represents a self-sufficient result, which can be of a special interest for some applications like that described in [9]. Further it can be considered in more details including the questions of tolerancing, mirrors manufacturing and testing, opto-mechanical design and assembling of the telescope.

**Funding.**

ERC(European Research Council) (H2020-ERC-STG-2015 – 678777).

**Acknowledgment.**

The authors acknowledge the support of the European Research council through the H2020 - ERC-STG-2015 – 678777 ICARUS program.